\documentclass[12pt]{article}
\usepackage{graphicx}
\usepackage{epsf} 
\usepackage{amsfonts,latexsym}
\begin{document}
\title{On quantum phase crossovers in finite systems} 

\author{Clare Dunning\S \footnote{email: t.c.dunning@kent.ac.uk}, 
Katrina E. Hibberd\P \footnote{email: keh@maths.uq.edu.au} and 
Jon Links\P \footnote{email: jrl@maths.uq.edu.au} \\
\S  Institute of Mathematics, Statistics and Actuarial 
Science, \\ The University of Kent, U.K. \\
\P  Centre for Mathematical Physics, School of Physical Sciences, \\
The University of Queensland, Brisbane 4072, Australia.}

\maketitle  

\begin{abstract}

In this work we define a formal notion of a quantum
phase crossover for certain Bethe ansatz solvable models.  
The approach we adopt exploits an
exact mapping of the spectrum of a many-body integrable system, which
admits an exact Bethe ansatz solution, into the quasi-exactly solvable
spectrum of a one-body Schr\"odinger operator.  Bifurcations of the
minima for the potential of the Schr\"odinger operator determine the crossover
couplings. 
By considering the behaviour of particular
ground-state correlation functions, these may be identified as quantum
phase crossovers in the many-body integrable system with finite particle
number.  In this approach the existence of the quantum phase crossover is not dependent
on the existence of a thermodynamic limit, rendering applications to finite systems feasible. 
We study two examples of bosonic Hamiltonians which
admit second-order crossovers.
~~~

\end{abstract}


\def\aa{\alpha} 
\def\bb{\beta}
\def\a{\hat a}
\def\b{\hat b}
\def\d{\dagger}
\def\de{\delta} 
\def\e{\epsilon}
\def\ve{\varepsilon}
\def\g{\gamma}
\def\K{\kappa}
\def\ap{\approx}
\def\l{\lambda}
\def\o{\omega}
\def\t{\tilde{\tau}}
\def\s{\sigma}
\def\D{\Delta}
\def\L{\Lambda}
\def\T{{\cal T}}
\def\TT{{\tilde{\cal T}}}
\def\E{{\cal E}} 
\def\f{\overline{f}}
\def\q{\overline{q}}
\def\tp{\otimes}
\def\I{\mathcal{I}} 
\def\N{\mathcal{N}}
\def\H{\mathcal{H}}
\def\rar{\rightarrow}

\def\bea{\begin{eqnarray}}
\def\eea{\end{eqnarray}}
\def\ba{\begin{array}}
\def\ea{\end{array}}
\def\no{\nonumber}
\def\le{\langle}
\def\re{\rangle}
\def\l{\left}
\def\r{\right}
\def\o{\omega}
\def\d{\dagger}
\def\nn{\nonumber}
\def\j{{ {\cal J}}}
\def\n{{\hat n}}
\def\A{{\cal A}}
\def\TT{{\tilde {\cal T}}}
\newtheorem{theorem}{Theorem}
\newtheorem{definition}{Definition}

\section{Introduction}

Quantum phase transitions may occur in the ground state (i.e. at
zero temperature) of quantum systems as an interaction coupling is varied.
Such phase transitions may be thought of as being driven by quantum
fluctuations, in analogy with thermal fluctuations underpinning thermal
phase transitions. In fact in many cases there is a correspondence between
a quantum phase transition in $d$ dimensions and a thermal phase
transition in $(d+1)$ dimensions \cite{sachdev}.  One of the important
mathematical tools in the study of thermal phase transitions is the
thermodynamic limit, where the number of particles is taken to infinity.
In this limit phase transitions can be associated with discontinuities in
certain physical quantities derived from the free energy.

For finite (i.e. mesoscopic) systems we cannot appeal to this notion of discontinuities to characterise a sharp change between quantum phases, as the transitions are smooth. Nonetheless for many finite systems there is a sense of a {\it crossover} between different quantum phases, and it useful to be able to characterise such a crossover, beyond saying that it is what occurs in a finite system in cases where there is phase transition in the thermodynamic limit. Quantum phases in finite systems have previously been studied in \cite{iz} 
in the context of the Interacting Boson Model (IBM). There it was argued that quantum phase transitions can be
identified via ``shape transitions'' of an effective potential energy
surface, defined in terms of classical variables, which becomes exact in
the thermodynamic limit. Our goal here is to formulate an approach which entirely avoids
the use of the thermodynamic limit, for applications to
cases where taking this limit may not be
desired or justified. For example there are known models
where crossovers between
ground-state phases only exist in the finite case. One possible situation is that the boundaries in parameter space between 
such ground-state phases may merge together as the thermodynamic limit is approached. An example of this is seen in the classical field-theoretic analysis of the attractive non-linear Schr\"odinger equation, where the transition coupling between the uniform regime and the broken symmetry soliton regime occurs at a coupling which scales as $N^{-1}$ for large $N$, where $N$ is the number of particles
\cite{ksu}. In the thermodynamic limit the transition coupling degenerates to zero (free-field case). 
For the Dicke model there are several ``critical points'' which degenerate in the thermodynamic limit
\cite{bor}. Another problematic scenario arises if taking a particular limit of a coupling parameter does not commute with taking the thermodynamic limit, as is known to happen in the weak coupling regimes of the BCS model \cite{silv,dlz} and the Bose gas with delta function interactions \cite{murray}.
Moreover for some systems, such as the attractive case of the Bose gas, there are
added technical difficulties in defining the thermodynamic limit, as the ground-state energy per particle is not finite in this limit \cite{murray,jim}.

Motivated by the above considerations, below we give a characterisation for a quantum phase crossover (QPC) in a one-body system with an external potential, for which there is no notion of a thermodynamic limit. Our definition for a QPC is given in terms of properties of the corresponding classical system. We then illustrate how this result can be used to investigate QPCs in Bethe ansatz solvable many-body interacting systems with finite particle number, through a manner which avoids taking the thermodynamic limit. The key to this approach, as will be detailed below,  is to exploit the Bethe ansatz solution to perform an exact mapping from the spectrum of the many-body interacting system into the spectrum of a one-body system in a potential.

\section{Quantum phase crossovers} \label{sect_qpc}
\subsection{One-body systems} \label{1_body}
We start with the Schr\"odinger operator (SO) eigenvalue equation in one dimension
\begin{equation} 
-\frac{\partial^2 \psi_k}{\partial x^2} + V(x)\psi_k = E_k \psi_k \label{wave} 
\end{equation} 
where for the potential $V(x)$ it is assumed there is a bifurcation of the global minimum at $x_0$, such that we may consider the approximation  
$$V(x)\approx V_0-2V_1(x-x_0)^2+V_2(x-x_0)^4$$
with $V_2>0$. A classical treatment of the above is equivalent to Landau theory, as used in the study of thermodynamic phase transitions. This approach has been previously discussed in \cite{iz}. The classical ground-state energy $\tilde{E}_0$ is given, to leading order in $V_1$, by 
\begin{equation}
\tilde{E}_0\sim \left\{\begin{array}{rr} V_0~~~~~~~~  & {\rm for ~}V_1<0\,\, \\ 
V_0-{V_1^2}/{V_2} & ~~~~~{\rm for ~}V_1>0. \end{array} \right.
\end{equation}   
Using this result as an approximation to the ground-state energy for the quantum system, we can appeal to the 
Hellmann--Feynman theorem to approximate the ground-state position fluctuations:
\begin{eqnarray*}
\left<(x-x_0)^2\right>=-\frac{1}{2}\frac{\partial \tilde{E}_0}{\partial V_1} \sim 
\left\{\begin{array}{rr} 0~~~~~  & {\rm for ~}V_1<0\,\, \\ 
{V_1}/{V_2} & ~~~~~{\rm for ~}V_1>0. \end{array} \right.
\end{eqnarray*}
In a full quantum analysis we must expect that quantum fluctuations will smooth out the discontinuity in the derivative of 
$\left<(x-x_0)^2\right>$ (and in particular $\left<(x-x_0)^2\right> >0$ for all $V_1$), so there is no quantum phase transition in the traditional sense. Nonetheless, the above classical analysis indicates that around the crossover coupling $V_1=0$ we should expect a sharp change in the behaviour of 
$\left<(x-x_0)^2\right>$ indicative of a crossover between a localised state and a Schr\"odinger cat state. Therefore we formally define a       
QPC for a one-dimensional SO as follows: 
\begin{definition}
Consider the SO equation  (\ref{wave}) where the potential $V(x)$ smoothly depends on a dimensionless coupling parameter $\gamma$. Treating (\ref{wave}) as a classical problem, approximate the ground-state energy as the minimum of the potential via 
$\displaystyle \tilde{E}_0=\min_{x\in\mathbb{R}} V(x).$   
If $m$ is the smallest integer for which 
${\partial^m \tilde{E}_0}/{\partial \gamma^m}$ is discontinuous at some coupling $\gamma_c$, we say there is an $m$th-order QPC of the quantum system at $\gamma_c$. 
\end{definition}
Of interest to the examples considered below are second-order crossovers, for which $\partial^2 \tilde{E}_0/\partial\gamma^2$
is discontinuous. 

\subsection{Finite many-body systems}

In the thermodynamic limit there are known examples for which there is a
one-to-one correspondence, established via the Bethe ansatz, between the
spectrum of an integrable model (IM) and the spectrum of a SO \cite{dt,clare,vladimir}.  
For IMs acting on
finite-dimensional Hilbert spaces an analogous scenario can hold, where
the spectrum of the IM maps into the spectrum of a SO, the so-called
quasi-exactly solvable (QES) sector \cite{quasi,uz}. Such a mapping is
obviously not one-to-one, but typically the QES spectrum of the SO
corresponds to low lying energy levels. In certain cases, such as the
examples studied below, the mapping is faithful between the ground-state
energies of the IM and the SO. This fact provides the means to explore
QPCs in the IM.

Let $H=H(\gamma)$ denote
the Hamiltonian for some IM acting on a finite-dimensional Hilbert space
where $\gamma$ is a dimensionless coupling parameter. Let $\left|\Psi_j\right>$ denote the
eigenstates of $H$ with energy levels $\mathcal{E}_j$ and the convention
that $\mathcal{E}_0$ is the ground-state energy.  
Now assume that the energies $\mathcal{E}_j$ can be mapped to 
energies $E_k$ of a SO
such that for each $j$ there exists a $k$ satisfying 
\begin{equation}
\mathcal{E}_j=  \chi E_k
\label{map}
\end{equation}
 for some {\it positive} scale factor 
$\chi$ which is independent of $\gamma,\,j$ and $k$. 
Since the mapping of the spectrum of the IM to the corresponding 
SO is not one-to-one, it is important to assert that the mapping is 
faithful between the ground-state energies. In our examples, 
following the arguments in \cite{quasi,uz}, this will be guarranteed 
by the following oscillation theorem \cite{bs}:
\begin{theorem}
Consider a SO with locally bounded potential $V(x)$ satisfying 
$V(x)\rightarrow\infty$ as $|x|\rightarrow\infty$. 
Let $\psi_k,\, k=0,1,\dots\infty$ denote the eigenfunctions of the 
SO with eigenvalues $E_k$ respectively, ordered such that $E_j< E_k$ whenever 
$j<k$. Then $\psi_k$ has precisely $k$ (real) zeroes.
\end{theorem}   
A corollary of the theorem is that the ground state wavefunction 
of the SO has no zeroes.  

Now assume that the mapping (\ref{map}) does hold between the ground state energies. 
Defining the operator $\mathcal{A}=\partial H/\partial \gamma$ which acts on the Hilbert space of the IM, we again use the Hellmann-Feynman theorem to deduce that for the ground state
$$\left<\mathcal{A}\right>= \frac{\partial \mathcal{E}_0}{\partial \gamma} = \chi
\frac{\partial {E}_0}{\partial \gamma} \approx \chi \frac{\partial {\tilde{E}}_0}{\partial \gamma} .$$
Thus the behaviour of the ground-state correlation function $\left<\mathcal{A}\right>$ will display a sharp change if 
$ {\partial^2 {\tilde{E}}_0}/{\partial \gamma^2} $ is discontinuous. This leads to
\begin{definition}
Suppose the ground state energy $\mathcal{E}_0$ of an IM exactly maps to the ground state energy ${E}_0$ of a SO through (\ref{map}). If the SO exhibits a second-order QPC at some dimensionless coupling $\gamma_c$ as in Definition 1, then we say that the IM also exhibits a second-order QPC at 
$\gamma_c$.   
\end{definition}   
Having outlined the theroetical considerations, we now apply it to specific models.

\section{Examples}
\subsection{Atomic-molecular bosonic model} 
The first model we look at describes the interconversion of bosonic atomic and di-atomic molecular modes \cite{vardi}. 
The Hamiltonian is
\begin{equation}
H=\frac{\delta}{2}n_a +\frac{\Omega}{2}(a^\dagger a^\dagger b + b^\dagger a a) 
\label{abham}
\end{equation}  
where $a^\dagger$ and $b^\dagger$ denote the creation operators for atomic and molecular modes respectively and as usual $n_a=a^\dagger a,\,n_b=b^\dagger b$. The total particle number $N=n_a+2n_b$ is conserved. In addition the Hamiltonian is invariant under the transformation $(a,\,a^\dagger)\rightarrow(-a,\, -a^\dagger)$. Since the change $\Omega\rightarrow -\Omega$ is equivalent to the unitary transformation $(b,\,b^\dagger)\rightarrow(-b, -b^\dagger)$ we restrict to $\Omega>0$. 

The Hamiltonian admits an exact Bethe ansatz solution \cite{zlm}. The solution depends on a discrete variable $p$ which takes values 0 or 1, depending on whether the total particle number is even or odd. The exact solution gives the energy levels as 
\begin{equation}
E=\delta \left(M+\frac{p}{2}\right)+\Omega\sum_{j=1}^M v_j \label{abnrg}
\end{equation}
where the parameters $\{v_j\}$ are the roots of the Bethe ansatz equations
\begin{equation}
\frac{2p+1}{2v_j}-v_j-\gamma=\sum_{k\neq j}^M\frac{2}{v_k-v_j}~~~j=1,\dots,M. \label{abbae}
\end{equation}
Above, $\gamma={\delta}/{\Omega}$ is defined as the dimensionless coupling, the number of roots $M$ 
is related to the total particle number through $N=2M+p$, and the dimension of the Hilbert space is $M+1$. 
From the ground-state energy, one can use the Hellman--Feynman theorem to compute the ground-state correlations
$$\left<n_a\right>=2\frac{\partial \mathcal{E}_0}{\partial \delta}, 
~~~~~\theta=-\frac{\partial \mathcal{E}_0}{\partial \Omega} $$
where 
\begin{equation}
\theta=-\left<a^\dagger a^\dagger b + b^\dagger a a\right>/2
\label{cc}
\end{equation}
 is the coherence correlator. 

\begin{figure}
\begin{center}
\includegraphics[scale=0.4]{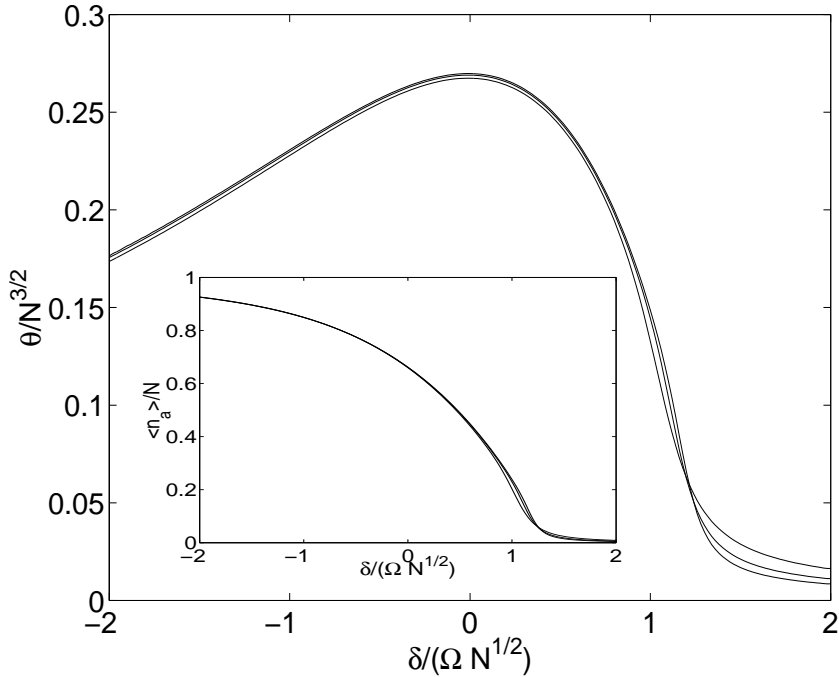}
\caption{\label{cf}
Behaviour of the coherence correlator (\ref{cc}) for the ground state of (\ref{abham}). 
The curves shown are for total particle 
number $N=20,30,40$. The inset shows $\left<n_a\right>/N$, the average fractional occupation of unbound atoms in the ground state. Below the  crossover coupling $\delta/(\Omega\sqrt{N})\approx 1.4$ the ground state is a coherent superposition of atomic and molecular states, while above the crossover coupling the ground state predominantly consists of molecular bosons.}
\end{center}
\end{figure}

Results of numerical analysis of the exact solution are shown in Fig. \ref{cf}, which has been taken from \cite{zlm}. 
Despite the very small number of particles, it can be deduced that below the crossover coupling $\delta/(\Omega\sqrt{N})\approx 1.4$ the ground state is a coherent superposition of atomic and molecular states, while above the crossover coupling the ground state predominantly consists of molecular bosons (see also Fig. 1 of \cite{kmcj}). The crossover sharpens with increasing $N$ but becomes singular in the thermodynamic limit $N\rightarrow\infty$, implying $\Omega\rightarrow 0$ and the eigenstates approach Fock states. However in this limit the ground-state energy per particle is not finite when $\delta/\Omega<0$ (see comments in the Conclusion).  We note that the qualitative features of $\left<n_a\right>/N$ shown in the inset are the same as those shown in Fig. 3 of \cite{iz} for the IBM. There it was argued the result could be interpreted as a second-order crossover, even for particle numbers of the order $10^1$. For (\ref{abham}) we will show that this same conclusion can be obtained from our approach described above.    

\begin{figure}
\begin{center}
\includegraphics[scale=0.8]{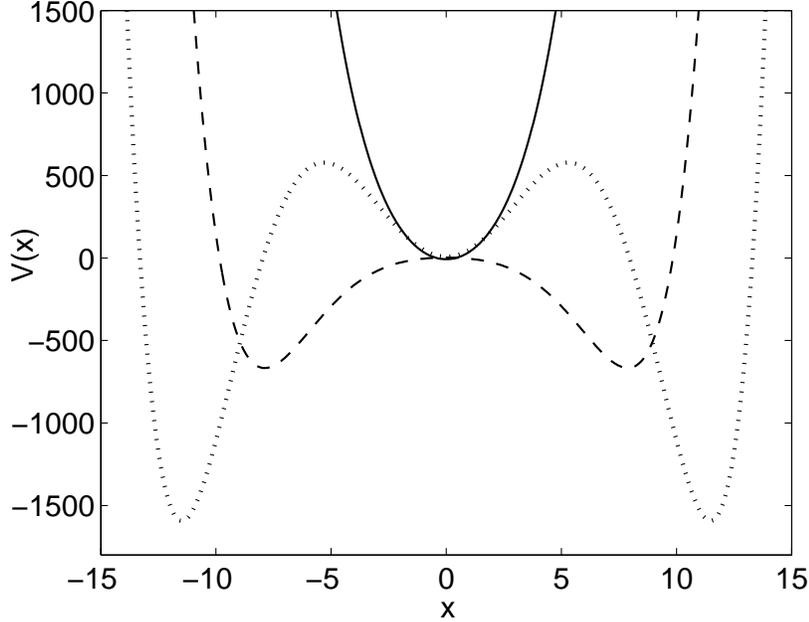}
\caption{\label{potential}
Generic behaviour of the sextic potential (\ref{abpot}) for different values of the dimensionless coupling $\gamma$, where the crossover coupling is given by $\gamma_c=\sqrt{2N+3}$. Here we take $N=100$. For $\gamma>\gamma_c$ 
(solid line - $\gamma=30$) there is a single minimum at $x=0$. For $-\gamma_c<\gamma<\gamma_c$ 
(dash line - $\gamma=-10$) or $\gamma<-\gamma_c$ (dot line - $\gamma=-30$) there are two global minima at 
$\pm x_0$ given by (\ref{xnought}).  
}
\end{center}
\end{figure}

To derive the value of the crossover coupling from the Bethe ansatz solution, we map the spectrum into that of a SO with a sextic potential. The procedure can be found in \cite{quasi}, so we simply give the result (see also \cite{uz}). 
For a given solution of (\ref{abbae}) with energy given by (\ref{abnrg}), we set  
\begin{equation}
\psi(x)= x^p\exp\left (-\frac{\gamma x^2}{8}-\frac{x^4}{64}\right)
\prod_{j=1}^M \left(\frac{x^2}{4}-v_j\right).   
\label{sexticstates}
\end{equation}
Then $\psi(x)$ satisfies (\ref{wave}) with $\chi=\Omega$  and potential 
\begin{equation}
V(x)=-\frac{\gamma}{4}+\frac{(\gamma^2 -3-2N)x^2}{16} 
+\frac{\gamma x^4}{32} +
\frac{x^6}{256}. 
\label{abpot}
\end{equation} 
The number of QES states (\ref{sexticstates}) is the dimension of the Hilbert space for (\ref{abham}), $M+1$. 
It is apparent that the states (\ref{sexticstates}) are even (odd) functions of $x$, with an even (odd) number of zeroes for $p=0$ ($p=1$) and at most $2M$ ($2M+1$) zeroes. In view of Theorem 1 no two states can have the same number of zeroes. For $p=0$ the QES states must be the states $\psi_k$ such that $k=0,2,4,\dots,2M$. Therefore the ground state of the SO with potential (\ref{abpot}) lies in the QES sector, and the mapping of the spectrum of (\ref{abham}) into that of the SO is faithful for the ground-state energies. For a system with odd particle number, the QES states correspond to the $k=1,3,\dots,2M+1$ states of the associated SO.
For this case one can alternatively consider the potential to be restricted to the half line $x\geq 0$ with the hard-wall
boundary condition $V(0)=\infty$, requiring $\psi(0)=0$. Then the QES states have an even number of zeroes in $(0,\infty)$ and 
correspond to the states $\psi_k$ such that $k=0,2,4,\dots,2M$, which includes the ground state. 
  
For $\gamma>0 $ and $(\gamma^2-3-2N)>0$ the potential (\ref{abpot}) attains its minimum at $x_0=0$, while for $(\gamma^2-3-2N)<0$
or $\gamma<0$ minima occur at 
\begin{equation}
x_0^2=\frac{4}{3}\sqrt{4{\gamma^2}-3(\gamma^2-3-2N)}-\frac{8\gamma}{3}. 
\label{xnought}
\end{equation} 
This analysis identifies the crossover coupling as $\gamma_c=\sqrt{2N+3}$. As $\gamma\rightarrow (\gamma_c)_-$ we find
$x_0\sim \pm (\sqrt{2N+3}-\gamma)^{1/2}$.
The above shows that for the potential (\ref{abpot}) there is a second-order QPC of the same universality class as the Landau (or mean-field) theory, which is also the same class for the second-order crossover of the IBM \cite{iz}. In the classical approximation the  behaviour of the correlation functions is found to be 
\begin{eqnarray}
 &&\theta-\theta_c \sim\gamma-\gamma_c,~~~~ \gamma\rightarrow (\gamma_c)_-  \nonumber \\
&&\theta -\theta_c\sim 0, ~~~~~~~~~~~ \gamma\rightarrow (\gamma_c)_+,   \nonumber
\end{eqnarray} 
with $\theta_c=0$. Similarly results hold for $\left<n_a\right>$. These are consistent with Fig. \ref{cf}. 

\subsection{Attractive two-site Bose--Hubbard model} Quantum tunneling of bosons, based on a two-mode approximation, 
can be described by the two-site Bose-Hubbard Hamiltonian \cite{leggett}:
\begin{equation}
H=-\frac{k}{8}(n_1-n_2)^2  
- \frac{{\mathcal E}}{2}(b^{\dag}_1 b_2 +b_2^{\dag} b_1)
\label{hamjo}
\end{equation}
where $b^{\dag}_j$, $j=1,2$ denote the single-particle creation 
operators associated with two bosonic modes   
and $n_1=b^{\dag}_1 b_1$ and $n_2=b^{\dag}_2 b_2$ 
are the corresponding number operators. The total particle number $N=n_1+n_2$ is conserved.
We only consider the attractive case for which $k>0$, and take $\E>0$. The case $\E<0$ can be obtained by the unitary transformation $(b_1,\,b^\dagger_1)\rightarrow(-b_1,\, -b_1^\dagger)$.

To derive the Bethe ansatz solution we follow the approach of \cite{uz,enolskii}. We start with
the Jordan-Schwinger realisation of the $su(2)$ algebra generators
\[ 
S^+= b^\dagger_1 b_2, \quad
S^-= b^\dagger_2 b_1, \quad
S^z=\frac{1}{2}(N_1-N_2) 
\] 
which satisfy the commutation relations
\begin{equation}
 [S^z, S^\pm]=\pm S^\pm,\qquad [S^+,S^-]=2S^z.
\label{su2}
\end{equation} 
The realisation is $(N+1)$-dimensional when the constraint of fixed particle number $N=N_1+N_2$ is imposed.
In terms of this realisation the Hamiltonian may be written 
\begin{equation}
H=-\frac{k}{2}(S^z)^2 -\frac{{\mathcal E}}{2}\left( S^++S^- \right).
\label{spinham}
\end{equation}
The same $(N+1)$-dimensional representation of $su(2)$ is given by the mapping to differential operators
\[ 
S^z= u\frac{{\rm d}}{{\rm d}u}-\frac{N}{2},\quad  
S^+=Nu-u^2\frac{{\rm d}}{{\rm d}u}, \quad
S^-=\frac{{\rm d}}{{\rm d}u} 
\]  
acting on the $(N+1)$-dimensional space of polynomials with basis $\{1,u,u^2,...,u^N\}$.  
We can then equivalently represent (\ref{spinham}) as the second-order differential operator
\begin{eqnarray}
H &=&-\frac{k}{2}\left(u^2\frac{{\rm d}^2}{{\rm d}u^2}+(1-N)u\frac{{\rm d}}{{\rm d}u}+\frac{N^2}{4}\right)
-\frac{{\mathcal E}}{2}\left(Nu+(1-u^2)\frac{{\rm d}}{{\rm d}u}\right) \nonumber \\ 
&=&-\frac{ku^2}{2} \frac{{\rm d}^2}{{\rm d}u^2} 
+\frac{1}{2}\left(k(N-1)u-{\mathcal E}(1-u^2)\right)  \frac{{\rm d}}{{\rm d}u}   
-\frac{kN^2}{8} -\frac{{\mathcal E}Nu}{2}.
\label{bhde}
\end{eqnarray}
Solving for the spectrum of the Hamiltonian (\ref{hamjo}) is then equivalent to solving the eigenvalue equation  
\begin{equation}
HQ=EQ
\label{eq:eigen}
\end{equation}   
where $H$ is given by (\ref{bhde}) and $Q(u)$ is a polynomial function of $u$ of order $N$. 
To obtain the Bethe ansatz solution, we first express $Q(u)$ in terms of its roots $\{v_j\}$:
\[
Q(u)=\prod_{j=1}^N(u-v_j). 
\] 
Evaluating (\ref{eq:eigen}) at $u=v_k$ for each $k$ leads to the set of Bethe ansatz equations 
\begin{eqnarray}
\frac{{\mathcal E}(1-v_k^2)+k(1-N)v_k}{k v^2_k}=\sum^N_{j\neq k}\frac{2}{v_j-v_k},\,\,\quad k=1,...,N.
\label{eq:bhd_bae}
\end{eqnarray}
Writing the asymptotic expansion $\displaystyle Q(u)\sim u^N - u^{N-1}\sum_{j=1}^N v_j$ and by  
 considering the terms of order $N$ in (\ref{eq:eigen}), the energy eigenvalues are found to be 
\begin{eqnarray}
E= -\frac{kN^2}{8} +\frac{{\mathcal E}}{2}\sum_{j=1}^N v_j.
\label{eq:bhd_nrg}
\end{eqnarray} 
We may exactly map the spectrum of (\ref{hamjo}) into that of the SO equation (\ref{wave}). Setting 
$\gamma={\mathcal E}/k$, $\chi=k/2$ and 
\begin{eqnarray*}
\psi(x)=  \exp(-\gamma\cosh(x))\prod_{j=1}^N\left(\exp\left(\frac{x}{2}\right)-v_j\exp\left(\frac{-x}{2}\right)\right) 
\end{eqnarray*}
then, as a result of (\ref{eq:eigen}), $\psi(x)$ satisfies (\ref{wave}) with the potential 
\begin{eqnarray}
V(x)&=&\gamma^2\sinh^2(x)-(N+1)\gamma\cosh(x)
\label{morse}
\end{eqnarray}
whenever the $\{v_j\}$ satisfy (\ref{eq:bhd_bae}).  
Using the same argument as in the previous example, it can be established that this mapping is 
faithful for the ground-state energies.

The potential (\ref{morse}) is a double Morse potential, which has previously been studied as a quasi-exactly solvable potential \cite{uz,kmmr}. It is a single well potential when $2\gamma>(N+1)$ with a mimimum at $x_0=0$, and a double well for $2\gamma<(N+1)$ with minima at 
$x_0=\pm\cosh^{-1}[(N+1)/2\gamma]$. We identify the crossover coupling as $\gamma_c=(N+1)/2$, 
and deduce that as $\gamma\rightarrow(\gamma_c)_-$ 
\begin{eqnarray*}
\pm x_0\sim ((N+1)/2-\gamma)^{1/2}
\end{eqnarray*} 
establishing that (\ref{hamjo}) is in the same universality class as (\ref{abham}). The coherence correlator 
in this instance, $\theta=-\langle b_1^\dagger b_2+b_2^\dagger b_1\rangle/2$, displays the classical critical behaviour 
 \begin{eqnarray}
 &&\theta-\theta_c \sim\gamma-\gamma_c,~~~~ \gamma\rightarrow (\gamma_c)_-  \nonumber \\
&&\theta -\theta_c\sim 0, ~~~~~~~~~~~ \gamma\rightarrow (\gamma_c)_+   \nonumber
\end{eqnarray} 
where $\theta_c=k\gamma_c$. 
We note that (\ref{morse}), with change of variable $x\rightarrow ix$, was previously derived in \cite{ads} as an exact {\it quantum phase model} for the repulsive case of (\ref{hamjo}). 

The QPC coupling $\gamma_c=(N+1)/2$ agrees, to leading order in $N$,
with the results of \cite{zoller}. In that work the QPC arises as the onset of broken symmetry in the mean-field analysis. The result is also in agreement with  \cite{pd}, where the QPC was identified by a numerical
analysis of wave function overlaps (cf. \cite{zanardi}) and ground-state entanglement measured
by the von Neumann entropy. In particular it was found that there was a
peak in the ground-state entanglement at the crossover coupling. This is
consistent with the claim of \cite{hines1} that a peak in the ground-state
entanglement occurs when there is a supercritical pitchfork bifurcation of
the global minimum of the Hamiltonian in the phase space of the
semi-classical limit. For (\ref{hamjo}), the semi-classical dynamics have
been studied in \cite{ks} and it is seen from the equations of motion that
such a bifurcation does occur at $\gamma_c$ (to leading order). It is interesting to note that in
contrast there is no peak in the entanglement at
$\gamma_c$ for the Hamiltonian (\ref{abham}) \cite{hines2}. For this latter case it has been shown 
in \cite{stfl} that the bifurcation occuring in the semi-classical phase space is not of the
supercritical pitchfork type.
 
\section{Conclusion}  
We have developed a technique for determining QPCs in certain
finite IMs admitting a Bethe ansatz solution, and applied it to the Hamiltonians (\ref{abham},\ref{hamjo}).
In both cases we identified a dimensionless crossover coupling $\gamma_c$,
whose role in defining the QPC is supported by the
behaviour of particular ground-state correlation functions. The
significance of the results is reflected by the fact that in both cases
$\gamma_c$ is $N$-dependent, which highlights the potential difficulties
in studying quantum phases in the thermodynamic limit. Furthermore, for the
Hamiltonian (\ref{abham}) the ground state energy per particle $\mathcal{E}_0/N$
scales as $N^{1/2}$ whenever $\gamma<0$ (as can be deduced from equations (\ref{abpot}) 
and (\ref{xnought})). For the Hamiltonian (\ref{hamjo}) the ground-state energy per particle at
$\gamma_c$ is given, in the classical approximation, by
$\mathcal{E}_0/N=kN(1+1/N)^2/4$. In both cases the ground-state energy per particle is not finite in the thermodynamic
limit.

Both models we have studied have only two degrees of freedom. In generic bosonic systems it is quite reasonable for a system to  have a low number of degrees of freedom, since the bosonic statistics allow for arbitrarily high particle number independent of the degrees of freedom. This situation is in stark contrast to fermionic systems where, due to the exclusion principle, increasing the number of particles is associated with increasing the number of degrees of freedom when taking the thermodynamic limit. 
A consequence of this difference between bosonic and fermionic systems is that for bosonic systems with small number of degrees of freedom, such as the two studied here, it is not possible to apply renormalisation group techniques to determine the nature of the ground-state phases. 

Finally, we comment on the feasability of extensions to other models.
Mappings from the spectrum of a many-body Hamiltonian to an SO can be
constructed in cases where there is a spectrum generating Lie algebra such
that the Hamiltonian commutes with the Casimir invariants. The $su(2)$
case has been discussed in \cite{kmmr}, where it was argued that 31
classes of quasi-exactly SOs exist.  Other examples can be found in
\cite{uz}. However it remains a challenge to determine the full extent of 
applicability of the methods we have introduced here.

\section*{Acknowledgement}
We gratefully 
acknowledge financial support from the Australian Research
Council through the Discovery Project DP0557949.
\vskip-.2in
 
\end{document}